\documentclass[twocolumn,aps,prd,unsortedaddress,%
superscriptaddress,a4paper,nofootinbib,nobalancelastpage,preprintnumbers,showpacs]{revtex4-1}
\usepackage{latexsym,amsbsy,amsmath}
  \usepackage[pdftex]{graphicx}
\usepackage{longtable}
 \DeclareGraphicsExtensions{.eps, .jpg,.png}
\renewcommand{\vec}[1]{\boldsymbol{#1}}
\begin{document}
\title{Stability of hexaquarks in the string limit of confinement}
\author{J.~Vijande}
\email{javier.vijande@uv.es}
\affiliation{Departamento de F\'{\i}sica At\'{o}mica, Molecular y Nuclear, Universidad de Valencia (UV)
and IFIC (UV-CSIC), Valencia, Spain.}
\author{A. Valcarce}
\email{valcarce@usal.es}
\affiliation{Departamento de F{\'\i}sica Fundamental,
Universidad de Salamanca, 37008 Salamanca, Spain}
\author{J.-M.~Richard}
\email{j-m.richard@ipnl.in2p3.fr}
\affiliation{Universit\'e de Lyon, Institut de Physique Nucl\'eaire de Lyon,
IN2P3-CNRS--UCBL,\\
4 rue Enrico Fermi, 69622  Villeurbanne, France}
\date{\today}
\begin{abstract}
 The stability of systems containing six quarks or antiquarks is studied 
within a simple string model inspired by the strong-coupling regime of quantum
chromodynamics and used previously for tetraquarks and pentaquarks. We discuss 
both six-quark $(q^6)$ and three-quark--three-antiquark $ (q^3\bar q{}^3)$
states.
The quarks are
assumed to be distinguishable and thus not submitted to antisymmetrization.
It is found that the  ground state of $(q^6)$ is stable against dissociation 
into two isolated baryons. 
For the case of $ (q^3\bar q{}^3)$, our results indicate the existence of a 
bound state very close to the threshold. The investigations are extended to
$(q^3Q^3)$ and $(Q^3\bar q^3)$ systems with two different constituent masses, and
their stability is discussed as a function of the mass ratio.
\end{abstract}

\maketitle

\section{Introduction}
The situation remains unclear and even confusing in the multiquark sector. 
Several experimental candidates have been announced and not confirmed. Also some
states with ordinary 
quantum numbers might be of multiquark nature or contain a large multiquark
component, but their interpretation is still controversial. For a review of the
experimental results, see, e.g., \cite{Nakamura:2010zzi}. It should be stressed,
however, that the recent experimental efforts have been devoted  mainly to
states with hidden heavy flavor, while other sectors have never been much
explored. 

On the theory side, there are some uncertainties on whether the models
describing 
ordinary mesons and baryons can be reliably extrapolated toward higher
configurations, and whether these tentative models do or do not lead to stable
multiquarks.  In particular, the dynamics of systems made either of six quarks,
$(q^6)$, or three quarks and three antiquarks, $(q^3\bar q{}^3)$, has been
discussed by 
several authors, for instance 
\cite{Jaffe:1976yi,Rosner:1985yh,Karl:1987cg,Fleck:1989ff,Oka:1988yq,
Shapiro:1978wi,Klempt:2002ap,Goldman:1987ma,Goldman:1989zj,Takeuchi:1990qj,
SilvestreBrac:1992yg,Cvetic:1980fh,Fredriksson:1981mh,Anselmino:1992vg,
Buchmann:1998mi,Chang:2004us,Ping:2006nc}, using  nuclear forces,
chromomagnetism, chiral
quark models, etc.

To focus on the role of confinement, we adopt here a simple string model 
inspired by \cite{Lenz:1985jk,Carlson:1991zt}. For mesons, it reduces to a
single linear potential, which can be scaled to $V_m=r_{12}$. For tetraquarks,
the so-called ``flip-flop'' interaction $V_t=\min(r_{13}+r_{24},r_{14}+r_{23})$
gives binding to equal-mass configurations $(qq\bar q\bar q)$ and to states with
two heavy quarks and two light antiquarks $(QQ\bar q\bar q)$
\cite{Vijande:2007ix}. Stable pentaquarks are also found in an extension of this
model \cite{Richard:2009rp}. The present article aims at studying hexaquark
states, both six-quark configurations or systems made of three quarks and three
antiquarks. 

Our string model is extremely crude, as it neglects altogether relativistic
effects, short range corrections, spin-dependent effects, etc.  Any
antisymmetrization is also disregarded, i.e., quarks are assumed of different
flavors, even if bearing equal masses.  There is no proliferation of multiquarks
in the experimental hadron spectrum, and antisymmetrization is certainly rather
effective in setting selection rules. However, before starting any detailed
calculation with a refined potential and a full account of Fermi statistics, we
wish to identify whether an improved picture of confinement favors the
occurrence of stable multiquarks.

In early multiquark calculations, indeed, 
the interquark potential was taken from the naive ansatz of additive terms with
color factors. Later, the flip-flop model was adopted and inserted in actual
few-body calculations. The good surprise in the tetraquark and pentaquark
cases \cite{Vijande:2007ix,Richard:2009rp} is that the flip-flop model gives
more attraction than the color-additive model, and thus suggests new scenarios
for multiquark binding.  Moreover, this model is supported by lattice QCD
\cite{Takahashi:2000te,Suganuma:2011ci}. This is an encouragement to extend the
study of stability in the six-quark sector.


This paper is organized as follows: 
in Sec.~\ref{se:models}, we present the linear string model which is adopted.
The methods to solve  the six-body problem are described in
Sec.~\ref{se:methods}. The results are presented in Sec.~\ref{se:results},
before some concluding remarks in Sec.~\ref{se:conclusions}.
\section{A simple string model}\label{se:models}
For mesons, the potential is taken to be the quark--antiquark
separation,
\begin{equation}\label{eq:model1-mes}
 V_m(1,2)=r_{12}~,
\end{equation}
the string tension being set to unity, to fix the energy scale. 

For a baryon  $(q^3)=\{1,2,3\}$, the potential is the now familiar $Y$-shape
potential
(see, e.g., \cite{Vijande:2007ix} for references to early papers on this
approach to the baryon dynamics)
\begin{equation}\label{eq:Ypot}
V_Y(1,2,3)=\min_{\ell} (r_{\ell 1}+r_{\ell 2}+r_{\ell 3})~.
\end{equation}
This potential can be estimated analytically by geometric considerations. If
$a$, $b$ and $c$ denote the sides of the triangle, namely $c=r_{12}$, etc.,  and
$\angle a$, etc., the opposite angles, the potential reads $V_Y(1,2,3)=b+c$ if
$\angle{a}>2\pi/3$, and permutations, and in the case where no angle exceeds
$2\pi/3$, (see, e.g., \cite{Takahashi:2000te})
\begin{eqnarray}\label{eq:Ypot1}
 &&V_Y(1,2,3)=\left[\frac{a^2+b^2+c^2+\Lambda^{1/2}}{2}\right]^{1/2}~,\\
&&\Lambda=3(a+b+c)(a+b-c)(a-b+c)(-a+b+c)~.\nonumber
\end{eqnarray}

\begin{figure}[htp]
 \centerline{%
\includegraphics[height=.12\textwidth]{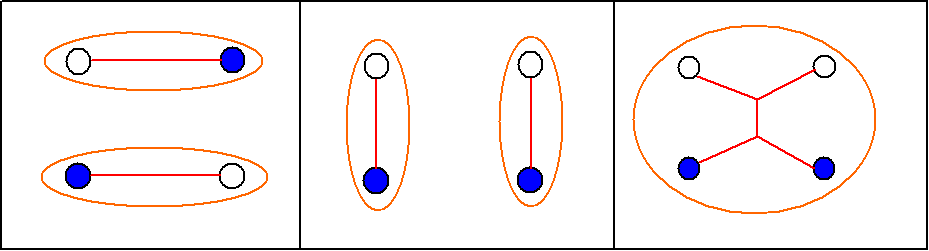}}
 \caption{Flip-flop interaction (left and center) and connected Steiner-tree diagram
(right) for the tetraquark. The
potential is in principle the minimum of the configurations, but it  is
largely dominated by the former ones.}
 \label{fig:mod1:tetra}
\end{figure}

For tetraquarks, the potential is taken to be minimum of the flip-flop
interaction
and the connected double-$Y$ diagram, both shown in
Fig.~\ref{fig:mod1:tetra}. It reads,
\begin{multline}\label{eq:model2-tet}
 V_4(1,2,3,4)=\min\bigg[r_{13}+r_{24},\, r_{14}+r_{23},\\
\min_{\{k,\ell\}} (r_{1k}+r_{2k}+r_{k\ell}+r_{\ell 3}+r_{\ell
4})\bigg]~.
\end{multline}

\begin{figure}[htp]
 \centerline{\includegraphics[height=.12\textwidth]{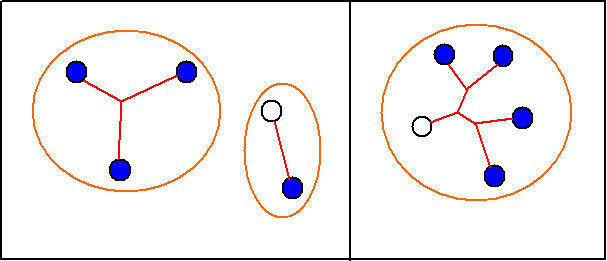}}
 \caption{Contributions to the pentaquark potential. Left: flip-flop. Right:
connected Steiner tree.}
 \label{fig:penta}
\end{figure}

For completeness, let us mention the pentaquark \cite{Richard:2009rp}, though it
will not enter any threshold nor sub-system in our study. The interaction is the
minimum of flip-flop terms, a meson and a baryon with all permutations, and of
connected Steiner trees, as shown in Fig.~~\ref{fig:penta}.

\begin{figure}[htp]
 \centerline{\includegraphics[width=.25\textwidth]{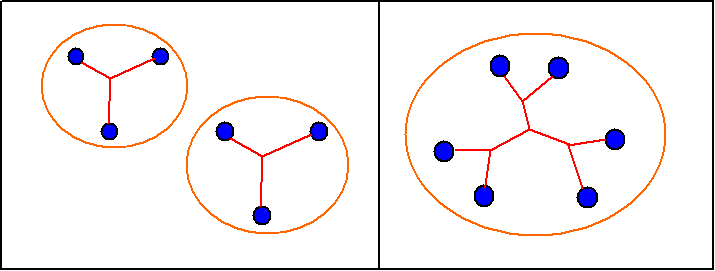}}
 \caption{Contributions to the $(q^6)$ potential.}
 \label{fig:dibaryon2}
\end{figure}

For the $(q^6)$ configurations, there are again two types of digrams: flip-flop
and connected Steiner tree, as shown in Fig.~\ref{fig:dibaryon2}. The potential
is minimized with respect to all permutations.                                  

\begin{figure}[!!!htp]
 \centerline{\includegraphics[height=.20\textwidth]{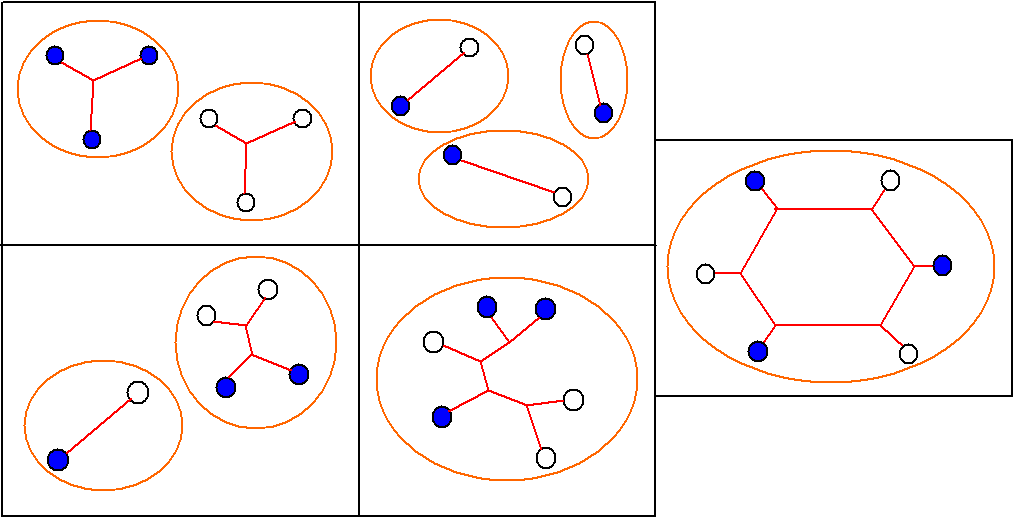}}
 \caption{Contributions to the $(q^3\bar q{}^3)$ potential.}
 \label{fig:baryonium2}
\end{figure}

Finally, for the $(q^3\bar q^3)$ states, there are several possibilities:
flip-flop with either a baryon and an antibaryon, or three mesons, or a meson
and a tetraquark, and also some connected diagrams with four or more junctions.
Examples are given in Fig.~\ref{fig:baryonium2}.

Now, the previous studies \cite{Vijande:2007ix} made on baryons, tetraquarks and
pentaquarks have shown that the dynamics is dominated by the flip-flop terms,
while the connected diagrams with $Y$-shape junctions play a minor role for
binding. Moreover, the
dynamics of baryon is qualitatively similar with a pair-wise potential $\sum
r_{ij}/2$ and the $Y$-shape model.\footnote{The main difference is that a
baryon bound by the $Y$-potential is slightly heavier than with $\sum
r_{ij}/2$. In a refined calculation of $(Q^3\bar q^3)$ with two different
masses, this would change the mass ratio at which there is a degeneracy of
baryon--antibaryon vs.\ mesonic thresholds, and perhaps influence the stability
of multiquarks in this region.} 

Hence for the ease of the
computations, we adopt from now on the following simplified interaction:
\begin{itemize}
 \item mesons: Eq.~\eqref{eq:model1-mes},
  \item baryons: the $\Delta$ interaction, %
\begin{equation}\label{eq:model1-bar}
 V_\Delta(1,2,3)=\frac{1}{2}(r_{12}+r_{23}+r_{31})~,
\end{equation}
 \item tetraquarks: the flip-flop terms, 
\begin{equation}\label{eq:model1-tet}
 V_t=\min(r_{13}+r_{24},\,r_{14}+r_{23})~,
\end{equation}
\item $(q^6)$: flip-flop with a $\Delta$ interaction for each baryon, with
suitable permutations,  see Fig.~\ref{fig:dibaryon1},
\begin{figure}[htp]
 \centerline{\includegraphics[width=.21\textwidth]{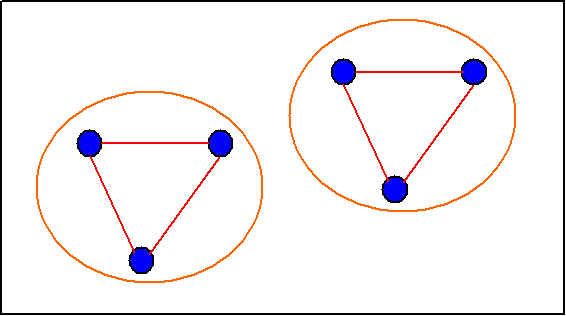}}
 \caption{Contributions to the dibaryon potential in the simplified model.}
 \label{fig:dibaryon1}
\end{figure}
\item $(q^3\bar q^3)$: only the flip-flop terms, with $\Delta$ for the baryon
and the antibaryon, and no double-$Y$ terms for tetraquark subsystems. See
Fig.~\ref{fig:baryonium1}.
The potential reads
\begin{multline}\label{eq:model1-baryonium}
 V_h(1,2,3,4,5,6)=\min\bigg[V_\Delta(1,2,3)+V_\Delta(4,5,6),\\
\min_{\{i,j,k\}}(r_{i4}+r_{j5}+r_{k6})\bigg]~.
\end{multline}
\begin{figure}[htp]
 \centerline{\includegraphics[width=.25\textwidth]{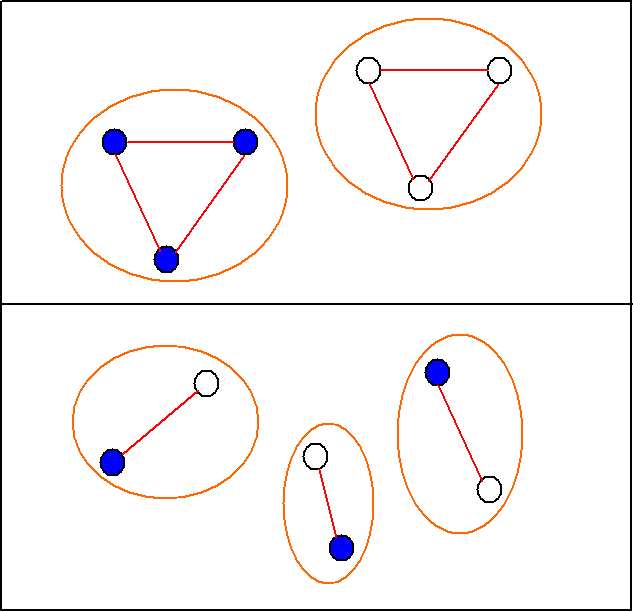}}
 \caption{Contributions to the $(q^3\bar q{}^3)$ potential in the simplified
model.}
 \label{fig:baryonium1}
\end{figure}
\end{itemize}
\section{Methods}\label{se:methods}
\subsection{Hyperspherical expansion}
The method of hyperspherical expansion is applicable for any set of
constituent masses, but we restrict its application to the case of equal
masses (but yet indistinguishable quarks).
One can describe the
relative motion with any standard set of Jacobi coordinates $\{\vec x_1,
\ldots,\vec x_5\}$, considered as a vector in a 15-dimensional space, with
spherical coordinates  $(r,\Omega)$. The potential $V(r,\Omega)$  is not exactly
isotropic, and the Schr\"odinger equation consists of an infinite set of coupled
equations 
for the radial reduced\footnote{a factor $r^7$ is included} partial waves
$u_L(r)$ with generalized angular momentum~$L$. A good (variational)
approximation consists of retaining only the $L=0$ (hyperscalar) component,
which obeys
$(m=\hbar=1)$%
\begin{equation}\label{eq:hyper-L0}
 -u''_0(r) + \frac{42}{r^2}u_0(r)+ V_{00} r u_0(r)=E_0 u_0(r)~,
\end{equation}
where  the projection
\begin{equation}\label{eq:hyper-proj}
V_{00}r=\left.\int V(r,\Omega)\mathrm{d}\Omega\right/ \int \mathrm{d}\Omega~,
\end{equation}
is computed numerically, unlike the case of a linear pairwise interaction, for which an analytic expression is available. This gives
\begin{equation}\label{eq:hyper-E0}
E_0(q^6)= 7.230\qquad  E_0(q^3\bar q{}^3)= 7.073~.
\end{equation}
For $(q^6)$, the potential is fully symmetric. As in the simpler three-body problem for baryons \cite{Richard:1992uk}, this implies that the next partial wave occurs only at $L=4$ and gives a very small correction. For $(q^3\bar q{}^3)$, there is a  $L=2$ contribution. One can solve the two coupled equations that generalize \eqref{eq:hyper-L0}, and get
\begin{equation}\label{eq:hyper-E2}
 E_2(q^3\bar q{}^3)=6.999~.
\end{equation}
\subsection{Correlated Gaussians}
The method of expansion over Gaussians has been used for cross-check in the
equal-mass 
case and extended to unequal constituent masses. 
This method is widely used in quantum chemistry and in few-body problems of 
nuclear physics \cite{Suzuki:1998bn,Hiyama:2003cu}, with some subtle variants
dealing with the most efficient manner of tuning  the parameters. 
In our case, it reduces to a  trial wave function  sought as
\begin{equation}\label{fbp:eq:gauss}
\Psi=\sum_{n=1}^N C_n\,\exp\left[-\sum_{i\geq j=1}^5 a_{n
ij}\,\vec{x}_i\cdot\vec{x}_j\right]~.
\end{equation}
Each individual term does not fulfill the constraints of permutation, parity,
etc., but, in principle, the proper symmetry requirements are restored in the
summation. 

\begin{figure}
 \centerline{\includegraphics[width=.38\textwidth]{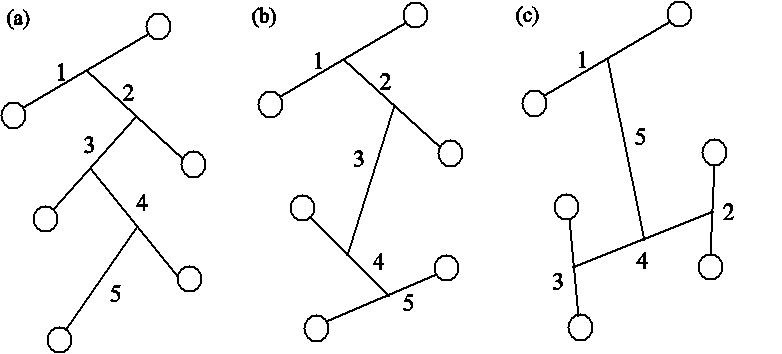}}
\caption{\label{fig-coor} Jacobi coordinates considered in this work.}
\end{figure}

As for the
Jacobi variables  $\vec x_i$, a simple and universal choice, (a), consists of 
\begin{equation}
\label{fbp:eq:jaco}
\begin{aligned}
\vec{x_1}&=\vec{r}_2-\vec{r}_1\\
\vec{x_2}&=\vec{r}_3-\frac{m_1}{m_{12}}\vec{r}_1-\frac{m_2}{m_{12}}\vec{r}_2\\
\vec{x_3}&=\vec{r}_4-\frac{m_1}{m_{123}}\vec{r}_1-\frac{m_2}{m_{123}}\vec{r}_2-\frac{m_3}{m_{123}}\vec{r}_3\\
\ldots&\ldots
\end{aligned}
\end{equation}
where $m_{1,2,..,n}=m_1+m_2+...+m_n$, as depicted in Fig.~\ref{fig-coor}(a). 
 As an alternative set, (b),
we can choose $\vec x_1$ and $\vec x_2$ to describe the internal motion in the
subset $\{1,2,3\}$, 
 $\vec x_4$ and $\vec x_5$ for $\{4,5,6\}$ and $\vec x_3$ for the relative
motion of 
the two clusters, see Fig.~\ref{fig-coor}(b). 
This generalizes the variables used in \cite{Vijande:2009zs} for the four-body
problem. We also  used a set (c), where the first coordinates describe the
two-body systems, and the last coordinates the relative motion of these
clusters, see Fig.~\ref{fig-coor}(c). 
In principle, the final result does not depend on the choice of Jacobi
coordinates. In practice, the number of terms, $N$ in Eq.~\eqref{fbp:eq:gauss}
is limited by the time spent in computing the matrix elements. With a finite
$N$, and incomplete restoration of symmetries, the results depend on the choice
of relative coordinates. 

For given $a_{nij}$, the linear parameters $C_n$ and the energy come from
solving a generalized eigenvalue problem. Then, the non-linear parameters $a_{nij}$ are
fitted to minimize the ground-state energy.
The matrix elements of the kinetic energy and normalization are known analytically. 
The multidimensional numerical integrals necessary to evaluate the potential
matrix 
elements have been solved using the CUBA package~\cite{2005CoPhC.168...78H}.
To guarantee the numerical accuracy of our results, several tests have been
performed. 
As for the convergence with  respect to the number $N$ of generalized
Gaussians in \eqref{fbp:eq:gauss},
we have pushed the calculation until the energy difference obtained using $N$
and $N-1$ 
Gaussians became smaller than the statistical uncertainty associated with the
Monte-Carlo integrations, of the order of 0.1\%. We have also used trial
wave functions in the different set of coordinates of Fig.~\ref{fig-coor}, looking
for the best and fast convergence.
Besides changing the Jacobi coordinates, we checked the scaling properties with 
respect to an overall factor applied to all masses, the virial theorem,
etc.  Our  results are consistent within 0.05\%.

For a given choice of Jacobi variables, using diagonal matrices, 
i.e., $a_{nij}=0$ for $i\neq j$, reduces the number of parameters. This means
that the internal orbital momenta are neglected. This approximation was made in 
\cite{Carlson:1991zt}, where the authors used a string potential similar to
ours,  and led them to conclude
that no six-quark bound states exist (even for the sole confinement potential, 
see Table I of Ref.\ \cite{Carlson:1991zt}). The effect of such approximation
over multiquark spectroscopy has been discussed in detail elsewhere 
\cite{Vijande:2007ix,Vijande:2003ki,Vijande:2009kj}.
 
\section{Results}\label{se:results}
\subsection{Thresholds}
In Table~\ref{thres-6q}, we compare the  threshold energies  for all 
possible decay channels.  For $(Q^3q^3)$, $(Qqq)+(QQq)$ is not shown, as it is
always above $(Q^3)+(q^3)$ \cite{Richard:1992uk}.

For moderate values of the quark-mass ratio $M/m$, the lowest threshold  
of $(Q^3\bar{q}^3)$ consists of 
a meson plus a tetraquark state, whose energy has been calculated 
in Ref.~\cite{Vijande:2007ix}. At higher values of  $M/m$ (not shown), the
lowest threshold becomes $(Q^3)+(\bar q^3)$.
\begin{center}
\begin{table}[t]
\caption{\label{thres-6q} $(Q^3q^3)$ and  $(Q^3\bar q{}^3)$ thresholds, 
as a function of the mass ratio
$M/m$, with the light quark mass set to $m=1$.}
\begin{ruledtabular}
\begin{tabular}{c|ccc}
$M$		& $(q^3)+(Q^3)$	&$(Q\bar q)^3$ 	&$(Q\bar q)+(Q^2\bar q^2)$ \\
\hline
1		& 7.728				& 7.011			&
6.981  \\
2		& 6.929				& 6.372			& 6.335	
 \\ 
3		& 6.543				&6.126			& 6.079	
\\
4		& 6.298				& 5.997			& 5.940	
\\
5		& 6.123				&5.916			& 5.852	
\\
\end{tabular}
\end{ruledtabular}
\end{table}
\end{center}

\subsection{Hexaquark energies}
The results are shown in Table \ref{results}. 
They correspond to three terms in the Gaussian expansion \eqref{fbp:eq:gauss}
using either the sets (a) or (b) of coordinates.

\begin{table}[t]
\caption{\label{results} $(Q^3q^3)$ and $(Q^3\bar q{}^3)$ variational energies
$E$, compared to their threshold energy $T$, as a function of the mass ratio
$M/m$, with the light quark mass set to $m=1$. $\Delta=E-T$ is the energy
diffference.}
\begin{ruledtabular}
\begin{tabular}{c|ccc}
$M$		& $E(Q^3q^3)$	& $T(Q^3q^3)$	& $\Delta(Q^3q^3)$	\\
\hline
1		& 7.237		& 7.728		& -0.491\\	
2		& 6.524		& 6.929		& -0.405\\	
3		& 6.209		& 6.543		& -0.334\\	
4		& 6.014		& 6.298		& -0.294\\	
5		& 5.890		& 6.123		& -0.233\\	
\hline
$M$& $E(Q^3\bar q{}^3)$	& $T(Q^3\bar q{}^3)$	& $\Delta(Q^3\bar
q{}^3)$\\
\hline
1 & 6.981		& 6.981		& +0.000	\\
2 & 6.314		& 6.335		& -0.021	\\
3 & 6.030		& 6.079		& -0.049	\\
4 & 5.868		& 5.940		& -0.072	\\
5 & 5.762		& 5.852		& -0.090	\\
\end{tabular}
\end{ruledtabular}
\end{table}

The ``dibaryon'', $(Q^3q^3)$, has been studied 
in the range of quark-mass ratio $1\le M/m\le 5$. For $M=m=1$, 
the result agrees quite well with the hypercentral approximation
\eqref{eq:hyper-E0}.
The system is found stable against dissociation into two baryons.
However, 
the stability deteriorates when the mass ratio increases. The behavior is
reasonably linear and therefore the  limit where the system becomes unbound can
be estimated to be of the order of $M/m\approx8-10$. Such a mass ratio 
corresponds to an intermediate value between the charm-to-light  and the
bottom-to-light
mass ratios. Hence a triple-charm dibaryon is predicted but not a
triple-beauty one. But departing from a pure linear potential would modify the
value of the critical mass ratio.

In the case of $(Q^3\bar q{}^3)$, this is more delicate. For $M=m=1$, the
results in Table~\ref{results} improve the hyperspherical estimate 
\eqref{eq:hyper-E2} truncated at $L=2$. 
It suggests that for a fully converged variational calculation, there is a 
shallow bound state below the lowest threshold.
This means that the effective interaction between the $(q\bar q)$ mesons is
attractive. Not surprisingly for these bosonic systems, if the dimer is bound,
the trimer is also bound, and a system $(q^k\bar{q}^k)$ even better for $k>3$,
though the neglect of any antimmetrization becomes less and less realistic for
$k>3$.  Also, for $k=3$, increasing the mass $M$ in the quark sector does not
modify much the effective interaction among $(q\bar q)$ mesons, 
and heavier mesons experience deeper binding. 
To check the existence of this bound state in our model, we repeated the 
calculation of the equal-mass case ($M=1$) with the set of coordinates (c). The
convergence turns out much faster. We got an energy $E\simeq6.860$ which
demonstrates a deeper binding.

Now, as $M/m$ further increases, the $(Q^3)+(\bar q{}^3)$ threshold will 
become degenerate with the lowest mesonic threshold. This will favor binding, as
the six-body wave function will contain two different decompositions into
clusters with relative motion that will interfere to improve binding.  However,
for even larger values of the mass ratio $M/m$, no multiquark configuration can
acquire enough binding to compete with the compact $(Q^3)$,  and the system
becomes unstable against rearrangement into $(Q^3)+(\bar q^3)$. Perhaps, some
metastability could be observed with respect to some higher threshold.
Similarly, models can be elaborated for hidden-charm or hidden-beauty resonances,
involving four-quark configurations that are in principle unstable against
$(Q\overline Q)+(q\bar q)$ but are mostly coupled to $(Q\bar q)+
\mathrm{c.c.}$.

Note that the six-quark energies shown in Table \ref{results} correspond to
rather 
different wave functions. For weak binding, the wave function is mostly a
hadron--hadron or a three-hadron molecule. Here, the improvements of the model
could be seek as long-range nuclear forces. In the case of deep binding, we are
dealing with a compact quark compound, and short-range quark forces,
chromomagnetic terms, and quark antisymmetrization are required to make the
estimate more realistic.
To disentangle the molecular from the compact structure of these objects, one
could 
proceed as in  Ref.~\cite{Vijande:2009zs} for $(Q^2\bar q{}^2)$. Unfortunately,
the necessary extension of the formalism  is far beyond the scope of the present
study. 

We now discuss briefly the sensitivity to details of the model, restricting 
ourselves to the equal-mass case.  We replace the perimetric $\Delta$-interaction 
of baryons or antibaryons is by the minima $Y$-path. 
For $(q^6)$, the threshold is now at $T=8.200$ if each baryon is estimated 
in the hyperscalar approximation. The same $L=0$ truncation for the six-body
problem gives an energy $E=7.650$. This means that the relative amount of
binding is very similar for both $\Delta$ and $Y$ cases.
For $(q^3\bar q{}^3)$, this is more intricate. The threshold is not changed, 
as it is made of a meson and a tetraquark.  However, the six-body potential is
slightly increased when changing the $\Delta$ interaction by the $Y$-one, and
not surprisingly, the ground-state energy also moves up, but stability is
preserved.

\section{Conclusions}\label{se:conclusions}
Let us summarize and suggest some possible further studies.

\textsl{1.} The string model of confinement which combines flip-flop and
connected 
flux tubes of minimal length gives more attraction than the additive pairwise
model with color factors that was used in early multiquark calculations. The
stability properties observed for some tetrraquark and pentaquark configurations
is confirmed in both the six-quark and the three-quark--three-antiquark sectors.

\textsl{2.} This potential  is flavor independent. By changing the constituent 
masses in the kinetic-energy part of the Hamiltonian, one can modify the
binding. For tetraquarks, the message was clear: $(QQ\bar q\bar q)$ is more
stable when the mass ratio $M/m$ increases, while $(Q\overline{Q}q\bar{q})$
becomes unstable. Here, the binding energy decreases for $(Q^3 q^3)$, while for
$(Q^3 \bar q{}^3)$, it first increases and then decreases. In the limit of large
$N$, no six-body configuration can compete with the deep binding of $(Q^3)$ that
enters the lowest threshold.

\textsl{3.} This potential can be seen as a simplistic Born--Oppenheimer limit. 
When the quarks or antiquarks move, the gluon fluxes readjust immediately into 
a connected or disconnected configuration with minimal cumulated length.  Thus
the color part of the quark wave function  is modified freely, without any
antisymmetrization constraint. The model requires changes to deal with identical
quarks. 

\textsl{4.} In the 70s, bumps were seen in the antiproton cross-sections and in 
the inclusive spectrum of antiproton annihilation, such as $\bar p+p\to \gamma
X$ \cite{Montanet:1980te}, not confirmed by experiments using improved
low-energy antiproton beams.  Recently, some enhancements have been observed in
the baryon--antibaryon mass distribution of $B$-meson or charmonium decay
\cite{Zhu:2007wz}.
 Models have been worked out  with a baryon and an antibaryon interacting by
mesonic exchanges or with $(q^2-\bar{q}^2)$ quark structure
\cite{Klempt:2002ap}.  The main uncertainty in the light quark sector lies in
the role of annihilation. Our model, which does not include any annihilation,
predicts some binding in the equal mass case. For flavor-asymmetric
configurations $(Q^3\bar{q}^3)$, which are free of annihilation, the binding is 
improved for moderate values of the quark-mass ratio. When $Q$ becomes
very large, the system breaks into a baryon and an antibaryon.

\textsl{5.} Dibaryon states have been often claimed but never firmly confirmed. 
The experimental situation  remains somewhat open, as some of the most recent
studies have given positive signals \cite{Khrykin:2000gh,Bashkanov:2008ih}. See
however, \cite{Skorodko:2009ys,Dymov:2009nr}.  Our model suggests the
possibility of stable dibaryon  states with exotic flavor configurations.  Years
ago, Jaffe pointed out the possibility of coherences in the
chromomagnetic interaction, and estimated that the $H(ssuudd)$ could be bound by
about $150\;$MeV below the $\Lambda\Lambda$ threshold. However, he used the
limit of flavor SU(3) symmetry and took for $(ssuudd)$  the short-range
correlation  coefficients as for ordinary baryons. Further studies indicated
that SU(3) breaking is not favorable
\cite{Rosner:1985yh,Karl:1987cg,Fleck:1989ff}, and that, not surprisingly, in
the dilute $(ssuudd)$, the strength of chromomagnetic effect is reduced as
compared to ordinary \cite{Oka:1988yq,Oka:2000wj}. However, early quark model
calculations used the ansatz of pairwise interactions with color factors, 
$V\propto \sum \tilde\lambda_i.\tilde\lambda_j \,v(r_{ij})$.  The string
potential gives more attraction.  It is interesting that two recent lattice-QCD
calculations \cite{Inoue:2010es,Beane:2010hg} of the $H$ conclude to the
possibility of a loosely bound or  a resonance close to the threshold.

\textsl{6.} The model would deserve further
variational calculations, with a larger variety of constituent masses, giving 
the possibility of playing with the relative location of baryonic vs.\ mesonic
thresholds.

\textsl{7.} Our aim is to reformulate the interaction as an operator in color
space, of which the present model will be the Born--Oppenheimer limit. This,
and to study the role of antisymmetrization in this new framework.

\acknowledgments
This work has been partially funded by the Spanish Ministerio de
Educaci\'on y Ciencia and EU FEDER under Contracts No. FPA2010-21750
and AIC10-D-000503,
and by the Spanish Consolider-Ingenio 2010 Program CPAN (CSD2007-00042).
Some discussions with E.~Hiyama, M.~Oka and K.~Yazaki are gratefully acknowledged.
%
%
%
%

\end{document}